\documentclass[secnumarabic, graphics,floatfix, nofootinbib,tightenlines,nobibnotes, aps, prl, 12pt]{revtex4}
\usepackage{graphicx}
\usepackage[english]{babel}
\usepackage[utf8]{inputenc}
\usepackage{amsmath,amssymb}
\usepackage{amsfonts}
\usepackage{multirow}

\begin{document}
 \newcommand{\bq}{\begin{equation}}
 \newcommand{\eq}{\end{equation}}
 \newcommand{\bqn}{\begin{eqnarray}}
 \newcommand{\eqn}{\end{eqnarray}}
 \newcommand{\nb}{\nonumber}
 \newcommand{\lb}{\label}
\newcommand{\PRL}{Phys. Rev. Lett.}
\newcommand{\PL}{Phys. Lett.}
\newcommand{\PR}{Phys. Rev.}
\newcommand{\CQG}{Class. Quantum Grav.}

\title{A Matrix Method for Quasinormal Modes: Kerr and Kerr-Sen Black Holes}

\author{Kai Lin$^{1,2,3)}$}\email{lk314159@hotmail.com}
\author{Wei-Liang Qian$^{3,4)}$}\email{wlqian@usp.br}
\author{Alan B. Pavan$^{1)}$}\email{alan@unifei.edu.br}
\author{Elcio Abdalla$^{5)}$}\email{eabdalla@usp.br}

\affiliation{1) Universidade Federal de Itajub\'a, Instituto de F\'isica e Qu\'imica, Itajub\'a, MG, Brasil}
\affiliation{2) Department of Astronomy, China West Normal University, Nanchong, Sichuan 637002, China}
\affiliation{3) Escola de Engenharia de Lorena, Universidade de S\~ao Paulo, Lorena, SP, Brasil}
\affiliation{4) Faculadade de Engenharia de Guaratinguet\'a, Universidade Estadual Paulista, Guaratinguet\'a, SP, Brasil}
\affiliation{5) Instituto de F\'isica, Universidade de S\~ao Paulo, S\~ao Paulo, Brasil}

\begin{abstract}
In this letter, a matrix method is employed to study the scalar quasinormal modes of Kerr as well as Kerr-Sen black holes.
Discretization is applied to transfer the scalar perturbation equation into a matrix form eigenvalue problem, where the resulting radial and angular equations are derived by the method of separation of variables.
The eigenvalues, quasinormal frequencies $\omega$ and angular quantum numbers $\lambda$, are then obtained by numerically solving the resultant homogeneous matrix equation.
This work shows that the present approach is an accurate as well as efficient method for investigating quasinormal modes.

\end{abstract}

\maketitle
\newpage

Black holes constitute an intriguing topic in astrophysics and theoretical physics, where the gravitational force is so strong that nothing can escape from inside of its event horizon.
The study of the properties of black holes might lead to insightful perspectives on quantum gravity.
The observation of many astronomical phenomena, as, for instance, gravitational lensing, become more accessible when associated to very compact stellar objects such as black holes.
Among others, one of the most important tools in the study of  black holes is the analysis of its quasinormal mode (QNM) oscillations, which describe the late time dynamics of black holes or black hole binaries, and therefore provide valuable information on the inherent properties of the black hole spacetime as well as its stability.
Recently, such signal was observed in the LIGO's first direct detection of gravitational wave \cite{LIGO,ligo-th}.

Generally, the QNM problem can be reformulated in terms of a Schr\"odinger-type equation.
Due to mathematical difficulties, an exact analytic solution is not always attainable.
Therefore, semi-analytical approximate and numerical methods have been proposed to calculate the quasinormal frequency (QNF) \cite{QNM1,QNM2,QNM3,QNM4,QNM5}, for example, the P\"oschl-Teller potential method \cite{PT Method}, continued fractions method \cite{continued fractions Method1,continued fractions Method2}, the Horowitz-Hubeny method (HH) for anti-de Sitter spacetime \cite{HH Method}, the WKB approximation \cite{WKB Method1,WKB Method2,WKB Method3}, the finite difference method \cite{FD Method1,FD Method2,FD Method3,FD Method4,FD Method5,FD Method6} and the asymptotic iteration method \cite{asymptotic iteration Method1,asymptotic iteration Method2,asymptotic iteration Method3,asymptotic iteration Method4} among others \cite{other Method1,other Method2,other Method3}.

In this letter, we make use of a matrix method \cite{Matrix Method} to calculate the scalar QNF's for rotating Kerr and Kerr-Sen black hole spacetimes.
By using the method of separation of variables, the radial and angular parts of the linearized perturbation equation of the scalar fields are given by \cite{Kerr,KerrSen}
 \bqn
 \lb{1}
(1-u^2)\frac{d}{du}\left[(1-u^2)\frac{dS_{lm}(u)}{du}\right]-\left[m^2-(1-u^2)\left(\lambda+a^2\omega^2u^2\right)\right]S_{lm}(u)&=&0 ,\nb\\
\frac{\Delta(r)}{r^2}\frac{d}{dr}\left[\frac{\Delta(r)}{r^2}\frac{dR_{lm}(r)}{dr}\right]+\frac{\Delta(r)}{r^6}\left[r^2V(r)+2\Delta(r)-r\frac{d\Delta(r)}{dr}\right]R_{lm}(r)&=&0 ,
 \eqn
where
 \bqn
 \lb{2}
V(r)&=&\frac{1}{\Delta(r)}\left\{[a^2+(2b+r)r]^2\omega^2-4Mam\omega r+a^2m^2\right\}-(a^2\omega^2+\lambda) ,\nb\\
\Delta(r)&=&r(r+2b)-2Mr+a^2 .
 \eqn
Here, $a\in [0,\frac12]$ gives the angular momentum per unit mass.
When $b\not=0$, it is the Kerr-Sen black hole case, which reduces to the Kerr black hole spacetime at $b=0$.
For the case of Kerr black hole, in order to compare our results with those from the continuous fraction method, the mass of the black hole is taken to be $M=1/2$.
On the other hand, for the case of Kerr-Sen black hole, the mass $M=(2b+r_0+r_i)/2$ and angular momentum $a=\sqrt{r_ir_0}$ can be expressed in terms of the event horizon $r_0$ and the inner horizon $r_i$.
$m$ represents the magnetic quantum number and $u=\cos\theta\in [-1,1]$.
It is noted that we have $\lambda=L(L+1)$ in the Schwarzschild limit with $a=0$ where $L$ is the angular quantum number.
By making use of the boundary conditions, one defines
 \bqn
 \lb{3}
R_{lm}(r)&=&\left\{
  \begin{array} {ll}
  e^{i\omega r}(r-r_0)^{-i\frac{2Mr_0\omega-am}{r_0-r_i}}r^{i\left[(r_0+r_i)\omega+\frac{2Mr_0\omega-am}{r_0-r_i}\right]}\Psi_R(r)~&\text{non-extreme case}, \nb\\
  e^{i\omega r+i\frac{2M\omega-m}{r-r_0}r_0}(r-r_0)^{2im-4iM\omega}r^{2i\left[(2M+r_0)\omega-m\right]}\Psi_R(r)&\text{extreme case}, \nb\\
\end{array}\right.\nb\\
S_{lm}(u)&=&e^{a\omega
u}(1+u)^{\frac{|m|}{2}}(1-u)^{\frac{|m|}{2}}\Phi_S(u)
 \eqn
with $r_0=M-b+\sqrt{(M-b)^2-a^2}$ and $r_i=M-b-\sqrt{(M-b)^2-a^2}$ being the event horizon and the inner horizon of the black hole (for the Kerr black hole with $M=1/2$, we simply have $r_0=\frac{1}{2}+\sqrt{\frac14-a^2}$ and $r_i=\frac{1}{2}-\sqrt{\frac14-a^2}$), respectively, so that the wave functions $\Psi_R(r)$ and $\Phi_S(u)$ are both finite at the boundaries.

Eq.(\ref{1}) is a specific form of the Teukolsky equation \cite{teq}.
It is noted that in our calculations we do not assume the value of $\lambda$ by using its Schwarzschild limit, which might lead to inaccurate results.
In what follows, we apply the matrix method to evaluate the QNF $\omega$ as well as $\lambda$.

First, we discretize an unkown wave function $f(x)$ at $N$ points $\{f(x_i)\}$ with $i=1,2,\cdots,N$.
Once convergence is assured, one can expand the function to all the points $x_i$ in the vicinity of $a$ by using the Taylor series,
 \bqn
 \lb{4}
f(x_i)&=&f(a)+(x_i-a)f'(a)+\frac{1}{2}(x_i-a)^2f''(a)+\frac{1}{3!}(x_i-a)^3f'''(a)+\cdots .
 \eqn
Now, repeatedly carrying out the above expansions around all the points $x_k$ with $k=1,2,\cdots,i-1,i+1,\cdots,N$, one obtains
\cite{Taylor method,Matrix Method}
 \bqn
 \lb{5}
{\Delta F}=M D,
 \eqn
where
 \bqn
 \lb{6}
{\Delta F}=\left(
    f(x_1)-f(x_k),
    f(x_2)-f(x_k),
    \cdots
    f(x_{k-1})-f(x_k),
    f(x_{k+1})-f(x_k),
    \cdots
    f(x_N)-f(x_k)
\right)^T \nb
 \eqn

 \bqn
 \lb{7}
M= \left(
  \begin{array}{cccccc}
    x_1-x_k & \frac{(x_1-x_k)^2}{2} &\cdots & \frac{(x_1-x_k)^i}{i!} &\cdots & \frac{(x_1-x_k)^{N-1}}{(N-1)!} \\
    x_2-x_k & \frac{(x_2-x_k)^2}{2} &\cdots & \frac{(x_2-x_k)^i}{i!} &\cdots & \frac{(x_2-x_k)^{N-1}}{{(N-1)}!} \\
        \cdots & \cdots & \cdots & \cdots & \cdots &\cdots \\
    x_{k-1}-x_k & \frac{(x_{k-1}-x_k)^2}{2} &\cdots & \frac{(x_{k-1}-x_k)^i}{i!} &\cdots & \frac{(x_{k-1}-x_k)^{N-1}}{{(N-1)}!} \\
    x_{k+1}-x_k & \frac{(x_{k+1}-x_k)^2}{2} &\cdots & \frac{(x_{k+1}-x_k)^i}{i!} &\cdots & \frac{(x_{k+1}-x_k)^{N-1}}{{(N-1)}!} \\
        \cdots & \cdots & \cdots & \cdots & \cdots &\cdots \\
    x_N-x_k & \frac{(x_N-x_k)^2}{2} &\cdots & \frac{(x_N-x_k)^i}{i!} &\cdots & \frac{(x_N-x_k)^{N-1}}{{(N-1)}!} \\
  \end{array}
\right)
 \eqn

 \bqn
 \lb{8}
D= \left(
    f'(x_k),
    f''(x_k),
    \cdots
    f^{(i)}(x_k),
    \cdots
    f^{(N-1)}(x_k)
\right)^T
 \eqn
Using Cramer's rule, we express the derivatives in $D$ as
 \bqn
 \lb{9}
f'(x_k)= \det(M_1)/\det(M),\nb\\
f''(x_k)= \det(M_2)/\det(M),
 \eqn
where $M_i$ is the matrix formed by replacing the $i$-th column of $M$ by the column vector $\Delta F$.
Therefore, when applying to Eq.(\ref{1}), it can be written in terms of a matrix equation by employing the above relation.
Consequently, $\omega$ and $\lambda$ in Eq.(\ref{1}) can be evaluated by solving the resultant non-linear eigenvalue problem.

To obtain the desired boundary conditions for the eigenvalue problem, we further introduce the coordinate transformation from $r$ and $u$ to $x$ and $z$, and transform the wave functions from $\Psi_R$ and $\Phi_S$ to $\chi_R$ and $\chi_S$ as
 \bqn
 \lb{10}
\chi_R&=& x(x-1)\Psi_R,\nb\\
\chi_S&=& z(z-1)\Phi_S,\nb\\
x&=&1-\frac{r_0}{r}\nb\\
z&=&\frac{1+u}{2}
 \eqn
where the variable $x\in [0,1]$ and the wave functions vanish on the boundary.
The master field equations in terms of $\chi_R$ and $\chi_S$ are thus obtained by substituting Eq.(\ref{2}), Eq.(\ref{3}) and Eq.(\ref{10}) into Eq.(\ref{1}),
 \bqn
 \lb{11}
\tau_R(x)\chi_R''(x)+\lambda_R(x)\chi_R'(x)+s_R(x)\chi_R(x)&=&0 ,\nb\\
\tau_S(z)\chi_S''(z)+\lambda_S(z)\chi_S'(z)+s_S(z)\chi_S(z)&=&0 ,
 \eqn
with the boundary conditions
\bqn
\chi_R(0)=\chi_R(1)=\chi_S(0)=\chi_S(1)=0 \; . \lb{bcon}
\eqn

By taking into account Eq.(\ref{9}), Eq.(\ref{11}) can be converted into the matrix form
 \bqn
 \lb{12}
\bar{\cal M}_R{\cal F}_R=0,~~~\text{and}~~~ \bar{\cal M}_S{\cal
F}_S=0\; ,
 \eqn
where ${\cal F}_R=\left(\chi_R(x_1),\chi_R(x_2),\cdots,\chi_R(x_N)\right)^T$ and ${\cal F}_S=\left(\chi_S(z_1),\chi_S(z_2),\cdots,\chi_S(z_N)\right)^T$, while $\bar{\cal M}_R$ and $\bar{\cal M}_S$ are two matrices with eigenvalues $\omega$ and $\lambda$ to be determined.
The boundary conditions can be implemented by replacing the first and the last row of the matrices $\bar{\cal M}_R$ and $\bar{\cal M}_S$ by Eq.(\ref{bcon}), or equivalently, one redefines the matrices
 \bqn
 \lb{13}
\left({\cal M}_R\right)_{k,i}= \left\{
  \begin{array}{cc}
    \delta_{k,i},  &k = 1~\text{or}~ N, \\
    \left(\bar{\cal M}_R\right)_{k,i}, & k = 2,3,\cdots,N-1 \\
  \end{array}
\right., \nb
\eqn
and
\bqn
\left({\cal M}_S\right)_{k,i}= \left\{
  \begin{array}{cc}
    \delta_{k,i},  &k = 1~\text{or}~ N, \\
    \left(\bar{\cal M}_S\right)_{k,i}, & k = 2,3,\cdots,N-1, \\
  \end{array}
\right.,
 \eqn
and replaces Eq.(\ref{12}) by
 \bqn
 \lb{14}
 {\cal M}_R{\cal F}_R=0 ,~~~\text{and}~~~{\cal M}_S{\cal F}_S=0\; .
 \eqn
The homogeneous equation Eq.(\ref{14}) implies that the eigenvalues $\lambda$ and $\omega$ satisfy the relation
 \bqn
 \lb{15}
\det( {\cal M}_R)=0 , ~~~\text{and}~~~\det( {\cal M}_S)=0\; .
 \eqn
It is obvious that the precision of the present method depends on the order of the expansion, $N$.
Therefore, in principle, higher accuracy can be achieved by using larger $N$.

\begin{figure*}[h]
\includegraphics[width=7.5cm]{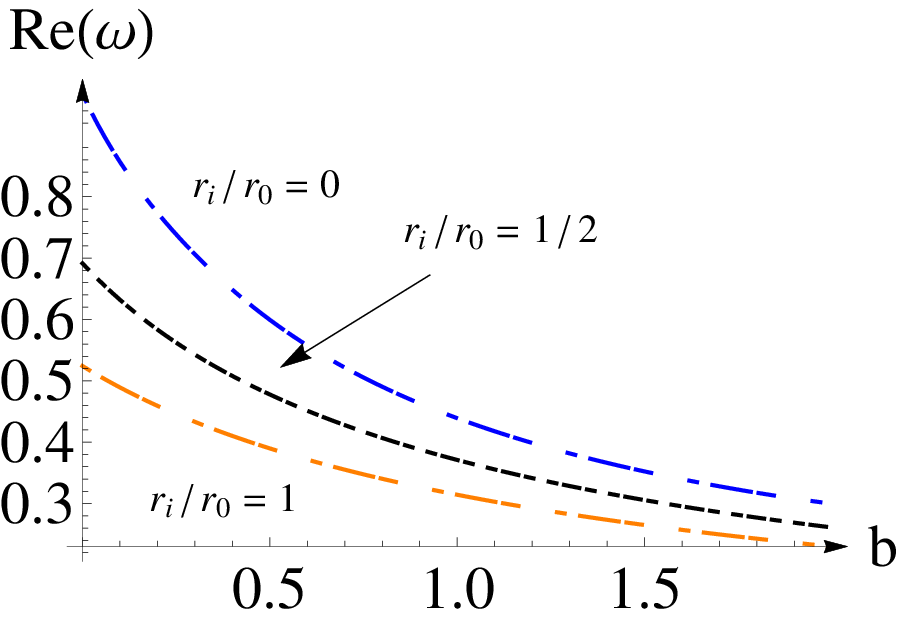}\includegraphics[width=7.5cm]{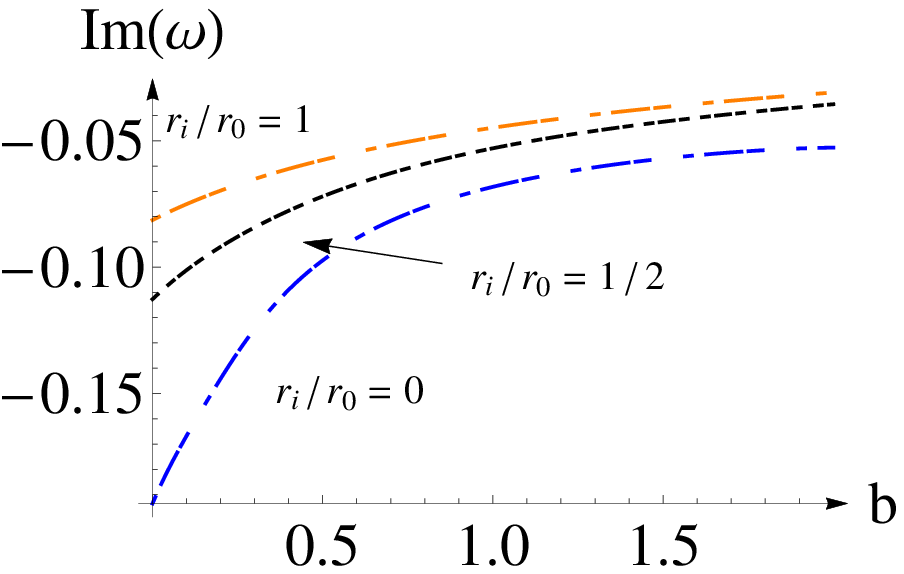}
\includegraphics[width=7.5cm]{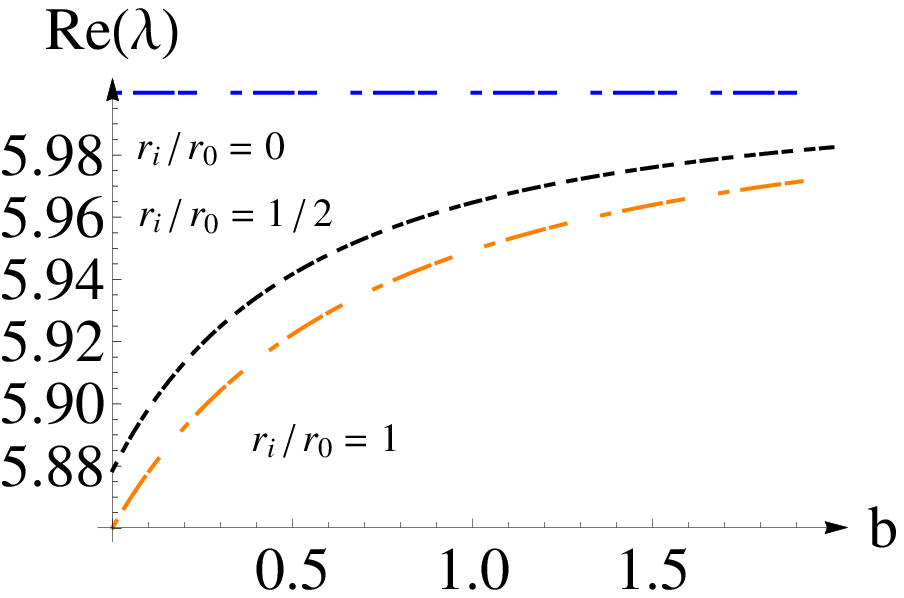}\includegraphics[width=7.5cm]{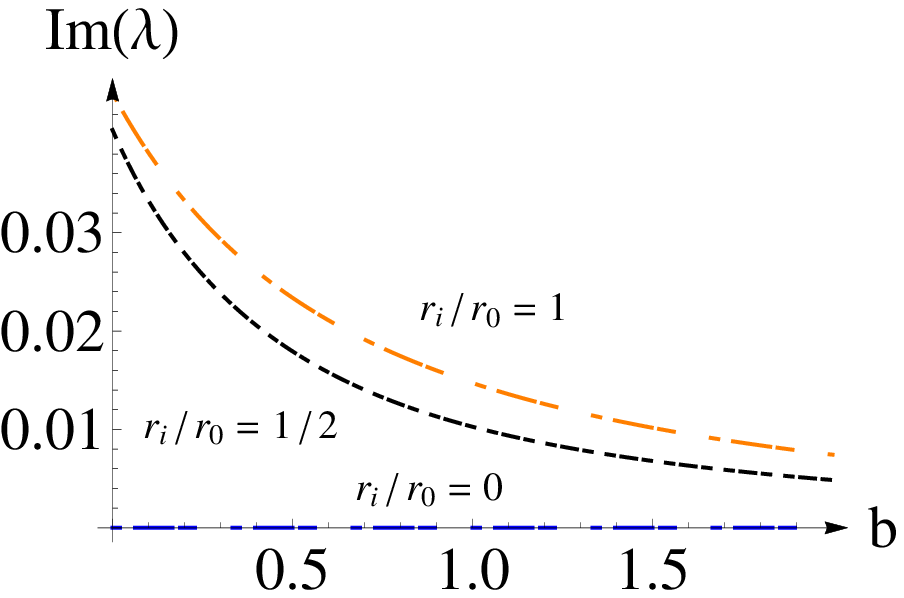}
\caption{The calculated $\omega$ and $\lambda$ of massless scalar field for the Kerr-Sen black hole with $m=0$, $r_0=1$ and $L=2$} \label{figb}
\end{figure*}

In Fig.\ref{figb}, the real and imaginary parts of $\omega$ and $\lambda$ of a massless scalar field for the Kerr-sen black hole are given as a function of $b$ by using the proposed matrix method.
The calculated curves are for different values of the inner horizon $r_i$ with $r_0=1$, $m=0$, and $L=2$.
As $b$ increases from $0$ to $1.5$, it is found that the real part of the QNF decreases for about 50\%, while the imaginary part of the QNF increases more sensitively and approaches zero.
Since $b$ measures the deviation of the theory from general relativity, if the QNF is not sensitive to the magnitude of $b$, it might imply certain indeterminacy in the black hole spacetime parameterization corresponding to the observed QNF.
Therefore, as pointed out recently \cite{roman}, such calculations might be relevant to the future LIGO and VIRGO data.

In order to study the precision of the proposed method, in Table I, we show the calculated $\omega$ and $\lambda$ for the Kerr black hole by the present matrix method, and they are compared to those obtained by the continued fractions method.
The present method is carried out by considering $N=18$ interpolation points.
For comparison purpose, the calculations by the continued fraction method are carried out up to the 130th order, and therefore the results can be viewed as exact.
The fourth column of the table gives the relative error of the present method.
It can be inferred from Table I that the deviations from the exact results are reasonably small, especially for the $L\ge 1$ cases.
By only considering a modest amount of terms in the discretization, the relative error is less than $10^{-5}$ in most cases.

\begin{figure*}[h]
\includegraphics[width=7.5cm]{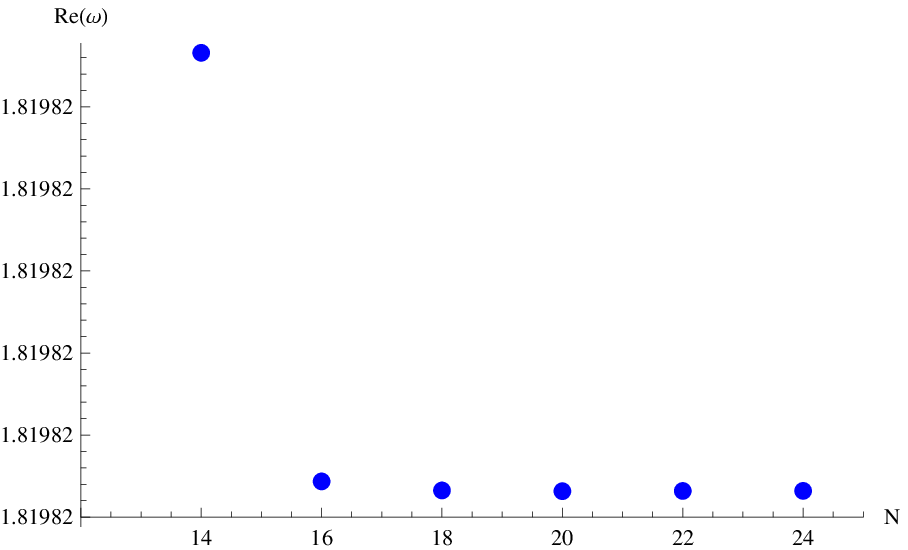}\includegraphics[width=7.5cm]{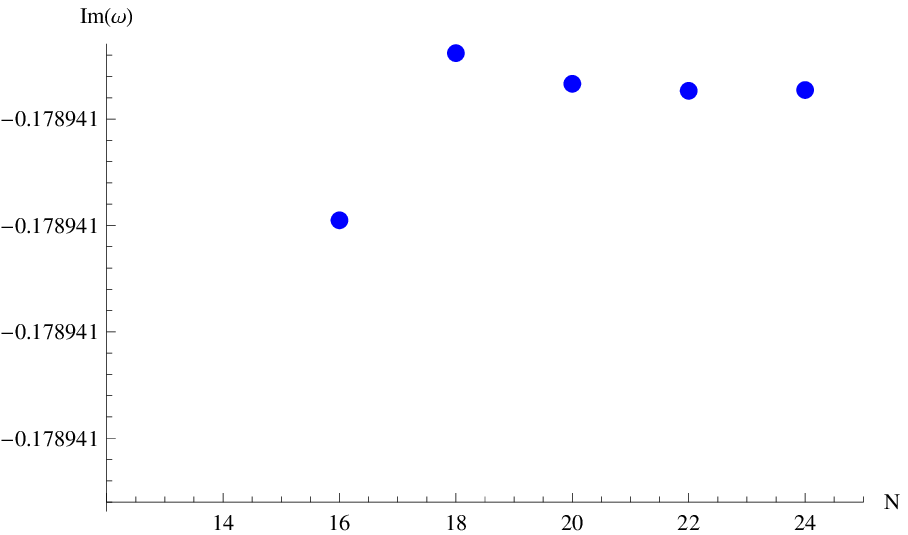}
\includegraphics[width=7.5cm]{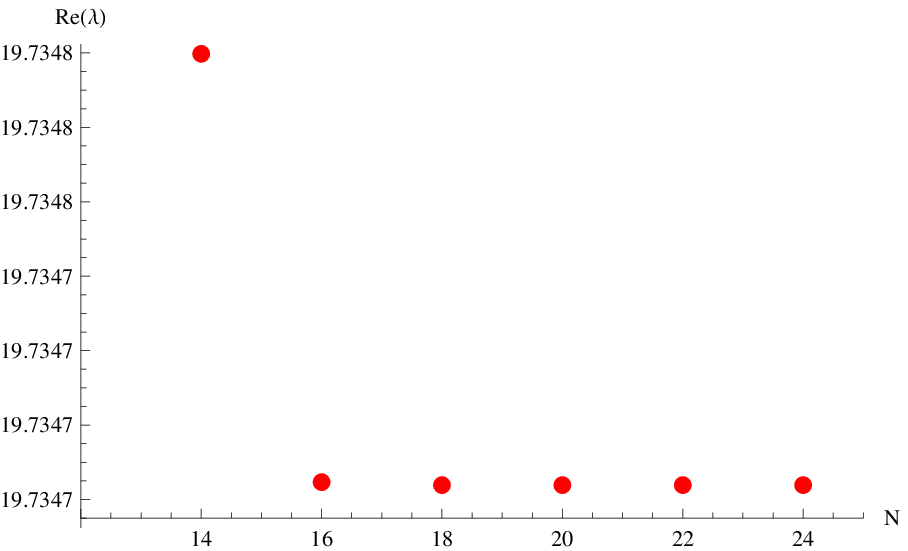}\includegraphics[width=7.5cm]{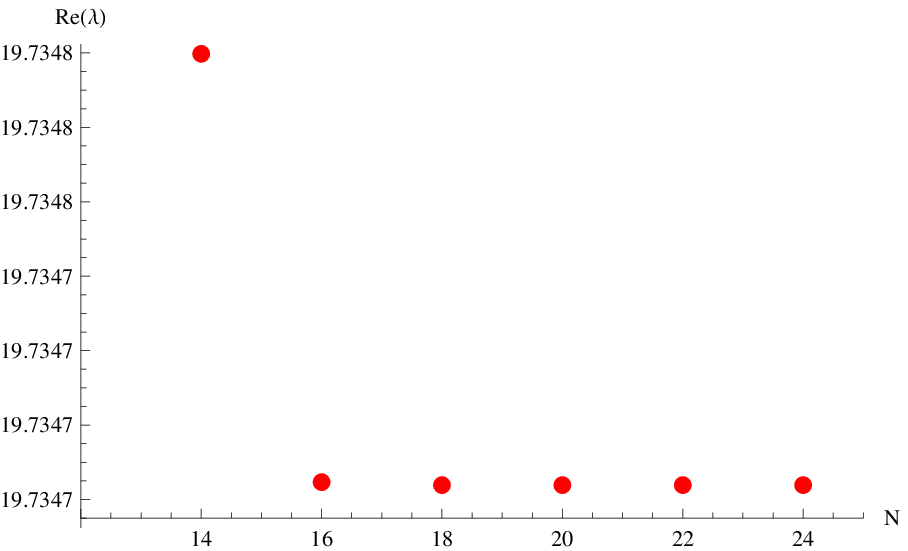}
\includegraphics[width=7.5cm]{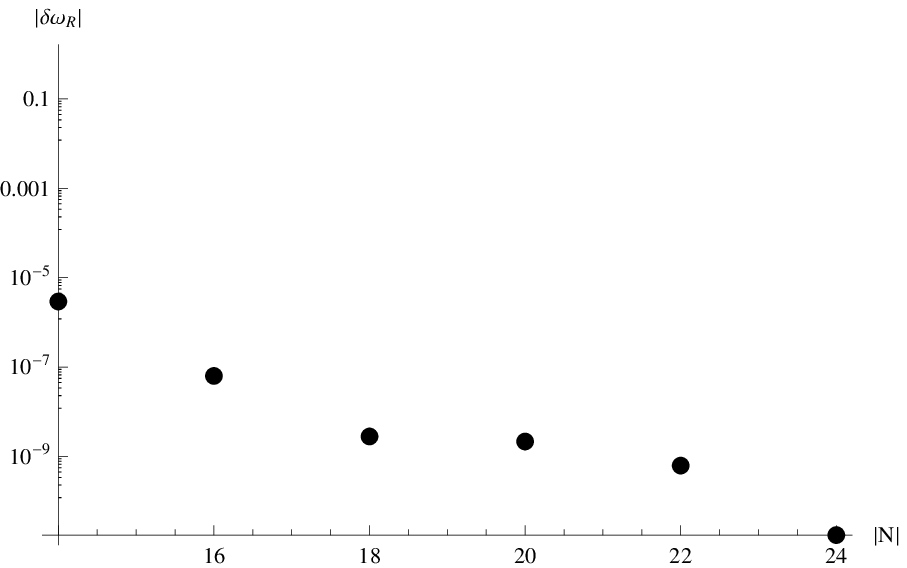}\includegraphics[width=7.5cm]{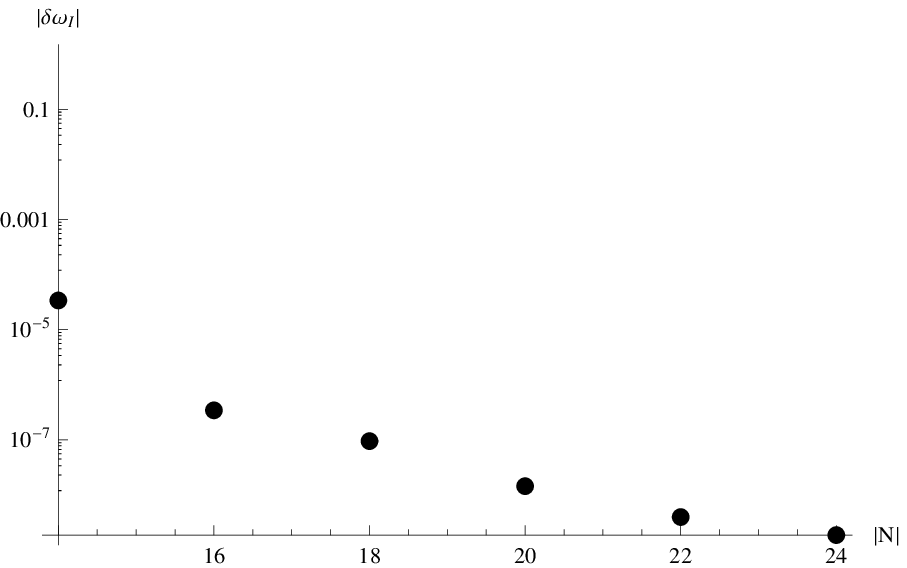}
\includegraphics[width=7.5cm]{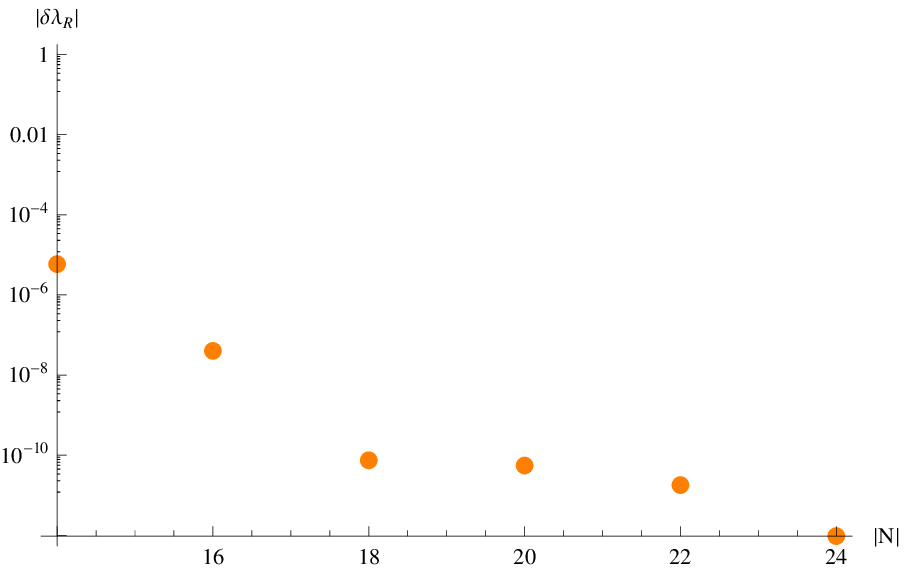}\includegraphics[width=7.5cm]{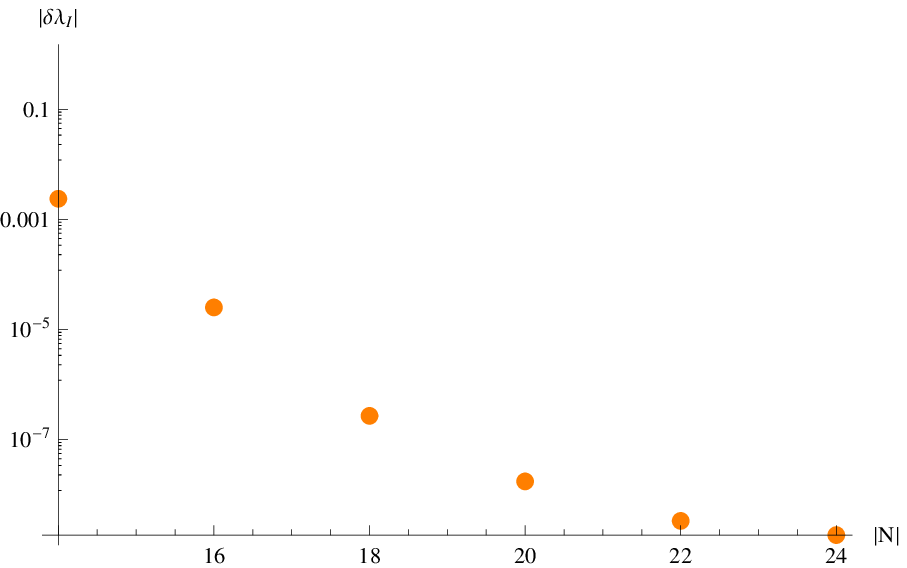}
\caption{The calculated values and their relative error of $\omega$ and $\lambda$ of massless scalar field for the Kerr black hole with $b=0$, $M=1/2$, $n=0$, $m=1$, $a=0.4$ and $L=4$.
The results are presented as a function of the number of interpolation points $N$.
The relative error is defined by $\delta A \equiv \frac{|A_c-A_e|}{|A_e|}$, where $A_c$ is the calculated result for quantity $A$ by the proposed matrix method, and $A_e$ is that by the continued fraction method (up to 130th order).}
\label{figerror}
\end{figure*}

In addition, we show in Fig.\ref{figerror} the calculated $\omega$ and $\lambda$, as well as their relative errors, as a function of the number of interpolation points $N$.
The calculations are carried out with $b=0$, $M=1/2$, $n=0$, $m=1$, $a=0.4$ and $L=4$.
It is found that the relative error becomes smaller than $10^{-5}$ once $N \ge 16$, and it continues to decrease with increasing $N$.
This convergence behavior indicates that the proposed method is accurate as well as stable.

Concerning the efficiency of the present method, since a major part of the method involves the solution of non-linear algebraic equations, it may take advantage of advanced algebraic equation solvers, such as {\it Mathematica} and {\it Matlab}. In general, we expect the proposed the method to have the same level of efficiency in comparison with several other methods based on non-linear algebraic equations, such as the continued fraction method, the HH method, etc.
However, a distinct feature of the matrix method is that a big chunk of analytic calculations associated with Eq.(\ref{9}) is independent of any specific metric, once $N$ is fixed.
In this regard, the method is flexible for dealing with different black hole background.
Moreover, the evaluation of Eq.(\ref{9}) can be carried out beforehand, which in turn saves the computation time.
Therefore, we come to the conclusion that the present method can be used as an efficient tool for the study of QNM.
We plan to apply the proposed method further to study more sophisticated black hole spacetimes in future investigations.

\section*{\bf Acknowledgements}

We gratefully acknowledge the financial support from Brazilian funding agencies
Funda\c{c}\~ao de Amparo \`a Pesquisa do Estado de S\~ao Paulo (FAPESP),
Conselho Nacional de Desenvolvimento Cient\'{\i}fico e Tecnol\'ogico (CNPq),
and Coordena\c{c}\~ao de Aperfei\c{c}oamento de Pessoal de N\'ivel Superior (CAPES),
as well as National Natural Science Foundation of China (NNSFC) under contract No.11573022 and 11375279.


\begin{table}[ht]
\caption{\label{TableI} The calculated $\omega$ and $\lambda$ of massless scalar field for the Kerr black hole ($b=0$) obtained by the present matrix method (MM) and by the continued fractions method (CFM).
The continued fraction method was carried out up to the 130th order and used as
The results are for the states with the principal quantum number $n=0$ and the mass of black hole $M=1/2$.
The relative errors are shown in the last column, the first value and the second value are the relative errors of real and imaginary parts of the QNF respectively.
The symbol $\cal O$ is used to indicate that the relative error is less than $10^{-5}$.}
\begin{tabular}{cccc}
         \hline
$(L,m,a)$~~&\text{MM with $N=18$}~~~~~~& \text{CFM with $N=130$}~~~~~~&\text{relative error}\\
        \hline
\{0,0,0\} &  $\omega$=0.221393 - 0.209814i&  $\omega$=0.22091 - 0.209791i&\{0.22\%,0.01\%\}  \\
          &  $\lambda$=0& $\lambda$=0&\{$\cal O$,$\cal O$\}  \\
\{0,0,0.2\} &  $\omega$=0.223892 - 0.206538i&  $\omega$=0.223398 - 0.206506i  &\{0.21\%,0.02\%\}\\
          &  $\lambda$=-0.0000993953 + 0.00123315i& $\lambda$=-0.0000966273 + 0.00123025i  &\{2.87\%,0.24\%\}\\
\{0,0,0.4\} &  $\omega$=0.229763 - 0.1914i&  $\omega$=0.229074 - 0.191404i &\{0.30\%,$\cal O$\} \\
          &  $\lambda$=-0.000858889 + 0.00469192i& $\lambda$=-0.00084195 + 0.00467792i  &\{2.01\%,0.30\%\}\\
        \hline
\{1,0,0\} &  $\omega$=0.585879 - 0.19532i&  $\omega$=0.585819 - 0.195523i  &\{0.01\%,-0.10\%\}\\
          &  $\lambda$=2& $\lambda$=2  &\{$\cal O$,$\cal O$\}\\
\{1,0,0.2\} &  $\omega$=0.592162 - 0.192525i&  $\omega$=0.592155 - 0.192526i  &\{$\cal O$,$\cal O$\}\\
          &  $\lambda$=1.99247 + 0.00547387i& $\lambda$=1.99247 + 0.00547382i  &\{$\cal O$,$\cal O$\}\\
\{1,0,0.4\} &  $\omega$=0.613394 - 0.18013i&  $\omega$=0.613388 - 0.180136i  &\{$\cal O$,$\cal O$\}\\
          &  $\lambda$=1.96698 + 0.0212409i& $\lambda$=1.96698 + 0.0212413i  &\{$\cal O$,$\cal O$\}\\
        \hline
\{2,0,0\} &  $\omega$=0.967288 - 0.193518i&  $\omega$=0.967284 - 0.193532i  &\{$\cal O$,$\cal O$\}\\
          &  $\lambda$=6& $\lambda$=6  &\{$\cal O$,$\cal O$\}\\
\{2,0,0.2\} &  $\omega$=0.97772 - 0.19093i&  $\omega$=0.97772 - 0.19093i  &\{$\cal O$,$\cal O$\}\\
          &  $\lambda$=5.98075 + 0.00781144i& $\lambda$=5.98075 + 0.00781144i  &\{$\cal O$,$\cal O$\}\\
\{2,0,0.4\} &  $\omega$=1.01425 - 0.179338i&  $\omega$=1.01425 - 0.179339i  &\{$\cal O$,$\cal O$\}\\
          &  $\lambda$=5.91671 + 0.0302983i& $\lambda$=5.91671 + 0.0302984i  &\{$\cal O$,$\cal O$\}\\
        \hline
\{1,1,0\} &  $\omega$=0.585879 - 0.19532i&  $\omega$=0.585872 - 0.19532i  &\{$\cal O$,$\cal O$\}\\
          &  $\lambda$=2& $\lambda$=2  &\{$\cal O$,$\cal O$\}\\
\{1,1,0.2\} &  $\omega$=0.663138 - 0.191587i&  $\omega$=0.663133 - 0.191583i  &\{$\cal O$,$\cal O$\}\\
          &  $\lambda$=1.99677 + 0.00203427i& $\lambda$=1.99677 + 0.00203422i  &\{$\cal O$,$\cal O$\}\\
\{1,1,0.4\} &  $\omega$=0.806547 - 0.166271i&  $\omega$=0.806545 - 0.166265i  &\{$\cal O$,$\cal O$\}\\
          &  $\lambda$=1.98003 + 0.008622i& $\lambda$=1.98003 + 0.00862165i  &\{$\cal O$,$\cal O$\}\\
        \hline
\{2,1,0\} &  $\omega$=0.967288 - 0.193518i&  $\omega$=0.967288 - 0.193518i  &\{$\cal O$,$\cal O$\}\\
          &  $\lambda$=6& $\lambda$=6  &\{$\cal O$,$\cal O$\}\\
\{2,1,0.2\} &  $\omega$=1.04428 - 0.190295i&  $\omega$=1.04428 - 0.190295i  &\{$\cal O$,$\cal O$\}\\
          &  $\lambda$=5.98192 + 0.00681854i& $\lambda$=5.98192 + 0.00681853i  &\{$\cal O$,$\cal O$\}\\
\{2,1,0.4\} &  $\omega$=1.18404 - 0.170251i&  $\omega$=1.18404 - 0.170251i &\{$\cal O$,$\cal O$\} \\
          &  $\lambda$=5.90568 + 0.027756i& $\lambda$=5.90568 + 0.0277559i  &\{$\cal O$,$\cal O$\}\\
        \hline
\{2,2,0\} &  $\omega$=0.967288 - 0.193518i&  $\omega$=0.967288 - 0.193518i  &\{$\cal O$,$\cal O$\}\\
          &  $\lambda$=6& $\lambda$=6 &\{$\cal O$,$\cal O$\} \\
\{2,2,0.2\} &  $\omega$=1.11929 - 0.189863i&  $\omega$=1.11929 - 0.189862i  &\{$\cal O$,$\cal O$\}\\
          &  $\lambda$=5.99304 + 0.00243193i& $\lambda$=5.99304 + 0.00243193i  &\{$\cal O$,$\cal O$\}\\
\{2,2,0.4\} &  $\omega$=1.41365 - 0.163041i&  $\omega$=1.41365 - 0.163041i  &\{$\cal O$,$\cal O$\}\\
          &  $\lambda$=5.95475 + 0.0106276i& $\lambda$=5.95475 + 0.0106275i &\{$\cal O$,$\cal O$\} \\
        \hline
\end{tabular}
\end{table}

\end{document}